\documentclass{jnmp}

\usepackage{amsmath}

\JNMPnumberwithin{equation}{section}

\newtheorem{theorem}{Theorem}

\theoremstyle{definition}

\newtheorem*{example}{Example} 

\newcommand{\arrow}{\rightarrow}
\newcommand{\bb}[1]{\mathbb{#1}}
\newcommand{\alg}{\mathfrak{g}}
\newcommand{\cea}{\overline{\mathfrak{g}}}
\newcommand{\Si}{\mathbb{S}^1}
\newcommand{\pr}{\partial}
\newcommand{\me}{\geqslant}

\newcommand{\bra}[1]{\left (#1\right )}
\newcommand{\brac}[1]{\left [#1\right ]}
\newcommand{\pobr}[1]{\left \{#1\right \}}
\newcommand{\eqreff}[2]{(\ref{#1}-\ref{#2})}
\newcommand{\pmatrx}[1]{\begin{pmatrix} #1 \end{pmatrix}}
\newcommand{\prh}{\hat{\partial}}
\newcommand{\Cs}{\mathcal{C}}
\newcommand{\alig}[1]{\left \{ \begin{aligned}#1\end{aligned}\right .}

\begin{document}

%
\renewcommand{\evenhead}{B M Szablikowski}
\renewcommand{\oddhead}{Gauge transformation and reciprocal link}

%
\thispagestyle{empty}

\FirstPageHead{*}{*}{20**}{\pageref{firstpage}--\pageref{lastpage}}{Article}

\copyrightnote{200*}{B M Szablikowski}

\Name{Gauge transformation and reciprocal link for
(2+1)-dimensional integrable field systems}

\label{firstpage}

\Author{B\l a\.zej M Szablikowski~$^\dag$}

\Address{$^\dag$ Institute of Physics, A. Mickiewicz University, Umultowska 85, 61-614 Pozna\'n, Poland\\
~~E-mail: bszablik@amu.edu.pl}

\Date{Received Month *, 200*; Revised Month *, 200*;
Accepted Month *, 200*}

\begin{abstract}
\noindent Appropriate restrictions of Lax operators which allows
to construction of (2+1)-dimensional integrable field systems,
coming from centrally extended algebra of pseudo-differential
operators, are reviewed. The gauge transformation and the
reciprocal link between three classes of Lax hierarchies are
established.\\
{\bf Scheduled for publication in Vol. 13 of 2006}
\end{abstract}

%

\section{Introduction}

It is well known that approach of the classical $R$-matrix
formalism to the specific infinite-dimensional Lie algebras can be
used for systematic construction of field and lattice integrable
dispersive systems (soliton systems) as well as dispersionless
integrable field systems (see \cite{BM}-\cite{RS-T-S2} and the
references there). The Lie algebra of pseudo-differential
operators (PDO) leads to the construction of (1+1)-dimensional
integrable soliton systems \cite{S-T-S,KO1,BS3}. Considering the
(1+1)-dimensional integrable hierarchies with infinitely many
fields one can extract from them closed equations for a single
field by the elimination of the remaining fields \cite{S,KD,OR}. This
method, the so-called Sato approach \cite{S}, leads to construction of (2+1)-dimensional integrable one-field equations:
Kadomtsev-Petviashvili (KP), modified Kadomtsev-Petviashvili (mKP)
and (2+1) Harry-Dym (HD). An analogous method with matrix
coefficients and dressing operators, the so-called matrix Sato
theory \cite{KO1}, permits a construction of (2+1)-dimensional
integrable evolution equations, with more number of fields, like
(2+1) AKNS. There is another more effective and systematic method
for construction of (2+1)-dimensional integrable systems, the
so-called central extension procedure \cite{RS-T-S1,RS-T-S2,PSA}.
The central extension approach to integrable field, lattice-field
and dispersionless systems was presented in \cite{BSP,BS} and
\cite{BS2}. In this paper an appropriate restrictions of Lax
operators coming from centrally extended PDO algebra, are systematically
considered.

In \cite{KO2,OR} a wide class of gauge, reciprocal,
B\"{a}cklund and auto-B\"{a}cklund transformations for
(1+1)-dimensional soliton systems and (2+1)-dimensional systems
like KP, mKP and HD, originating from the PDO Lie algebra, is
presented. Therefore, the investigation of such transformations
for (2+1)-dimensional systems, originating from the centrally
extended PDO algebra, seems interesting. In this paper the
relations between three classes of Lax hierarchies, coming from
the centrally extended PDO algebra, are constructed.

\section{$R$-matrix and the central extension approach}

The crucial point of the formalism is the observation that
integrable systems can be obtained from Lax equations. Let $\alg$
be a Lie algebra, equipped with the Lie bracket $[\cdot, \cdot]$.
A linear map $R:\alg \arrow \alg$, such that the bracket $[a,b]_R
:= [R a, b] + [a,R b]$ is a second Lie product on $\alg$, is
called the classical $R$-matrix. Let $R$ satisfy an Yang-Baxter
equation YB($\alpha$): $[R a, R b] - R [a, b]_R + \alpha [a,b] =
0$, which is a sufficient condition for $R$ to be an $R$-matrix.
Assume now that the Lie algebra $\alg$ depends effectively on an
independent parameter $y\in \Si$, which naturally generates the
corresponding current operator algebra $\cea =
\Cs^{\infty}(\Si,\alg)$. Then, invariants $A_n\in $, numbered by
$n$, of the Novikov-Lax equation, i.e.
\begin{equation}\label{novlax}
    \brac{A_n,L}+\bra{A_n}_y = 0\qquad \qquad L\in \cea,
\end{equation}
generate mutually commuting vector fields
\begin{equation}\label{eveq}
    L_{t_n} = \brac{R A_n, L}+ \bra{R A_n}_y .
\end{equation}
For a fixed $n$, the remaining systems are considered as its
symmetries. In this sense \eqref{eveq} represents a hierarchy of
integrable dynamical systems.

To construct the simplest $R$-structure let us assume that the Lie
algebra $\alg$ can be split into a direct sum of Lie subalgebras
$\alg_+$ and $\alg_-$, i.e. $\alg =\alg_+ \oplus \alg_-$, where
$[\alg_\pm,\alg_\pm]\subset \alg_\pm$. Denoting the projections
onto these subalgebras by $P_\pm$, we define the $R$-matrix as $ R
= P_+ - \frac{1}{2}$. Then, straightforward calculation shows that
$R$ solves YB($\frac{1}{4}$). Following the above scheme, we are
able to construct in a systematic way integrable systems once we
fix a Lie algebra.

\section{Centrally extended PDO Lie algebra}

Let us consider the associative algebra of pseudo-differential
operators
\begin{equation}\label{psdalg}
\cea = \pobr{L = \sum_{i\in \bb{Z}} u_i(x,y)\prh_x^i},
\end{equation} where the coefficients
$u_i(x,y)$ are dynamical fields being smooth functions and
$y\in \Si$ is an additional independent variable introduced by the
central extension procedure. We use a dash to distinguish the
pseudo-differential operators such $\prh_x$ and $\prh_x^{-1}$ from
derivatives $\pr_x$ and from formal integration symbols
$\pr_x^{-1}$. Multiplication of two operators in $\cea$ uses the
generalized Leibniz rule: $\prh_x^i u =\sum_{s\me 0} \binom{i}{s}
u_{sx} \prh_x^{i-s}$, where $\binom{i}{s} = (-1)^s
\binom{-i+s-1}{s}$ for $i<0$. Lie structure on $\cea$ is given by
the commutator $[A, B] = A B -B A$. The algebra \eqref{psdalg} can
be split into a direct sum of its Lie subalgebras $\cea_{\geq k} =
\{\sum_{i\geq k} u_i\prh_x^i\}$ and $\cea_{< k} = \{\sum_{i< k}
u_i\prh_x^i\}$ for $k = 0, 1, 2$. As a result, $R$-matrix is given
by $R= P_{\geq k} - \frac{1}{2}$. Then, the
hierarchy \eqref{eveq} takes the form
\begin{equation}\label{exev}
    L_{t_n} = \brac{\bra{A_n}_{\geq k},L - \prh_y}\qquad \qquad k = 0,1,2 .
\end{equation}
We are interested in extracting closed systems for a finite number
of fields. To obtain a consistent Lax equation, the Lax operator
$L$ has to span a proper subspace of the full algebra $\cea$, i.e.
the left and right-hand sides of expression \eqref{exev} have to
coincide. To construct any integrable equations for a given $L$ we
have to know how to solve \eqref{novlax}, i.e.
\begin{equation}\label{exnov}
    \brac{A_n, L-\prh_y} = 0 .
\end{equation}
It can be done by putting $A_n = \sum_{i\leq n} a_i \prh_x^i$ for
$n\geq k$. Then the function parameters $a_i$ are obtained from
\eqref{exnov} successively via the recurrent procedure. Note that
solutions are in the form of infinite series. In fact we need only
the finite parts $(A_{n})_{\geq k}$. We will look for the restrictions
of Lax operators $L$ given in the general form:
\begin{equation}\label{genl}
L = u_N\prh_x^N + u_{N-1} \prh_x^{N-1} + ... + u_0 +
\prh_x^{-1}u_{-1} + ... + \prh_x^{-m}u_{-m}.
\end{equation}

There is another important class of restrictions to the Lax
operators, which contain additional dynamical fields, the
so-called source terms $\psi_i$ and $\phi_i$ \cite{OS}, given by
\begin{equation}\label{exl}
L = l + \sum_i \psi_i \prh_x^{-1} \phi_i,
\end{equation}
where $l$ has an appropriate form as \eqref{genl}. For
pseudo-differential operator $A=\sum_i a_i\prh_x^i$ let
$[A]_0=a_0$, so $[\cdot]$ means the projection to the $0$-order.
The transposition operation of $A$ is defined as $A^\dag = \sum_i
(a_i\prh_x^i)^\dag = \sum_i (-1)^i \prh_x^i a_i$. Hence, from the
Lax equation \eqref{exev} and the equivalent transposed Lax
formulation
\begin{equation}\label{trexev}
L_{t_n}^\dag = -\brac{\bra{A_n}_{\geq k}^\dag, L^\dag +
\prh_y}\qquad \qquad k=0,1,2,
\end{equation}
it follows that $\psi_i$ and $\phi_i$ are eigenfunctions and
adjoint-eigenfunctions of the Lax hierarchy \eqref{exev}, i.e.
they satisfy
\begin{equation*}
\bra{\psi_i}_{t_n} = \brac{\bra{A_n}_{\me k}\psi_i}_0\qquad
\bra{\phi_i}_{t_n} = -\brac{\bra{A_n}_{\me k}^\dag \phi_i}_0 .
\end{equation*}

\paragraph{The case: $k=0$.}\hfill \\

The simplest appropriate Lax operators are given in the form
\begin{equation}\label{k0}
L = \prh_x^N + u_{N-2} \prh_x^{N-2} + u_{N-3} \prh_x^{N-3} + ... +
u_1 \prh_x + u_0\qquad \quad N\me 2.
\end{equation}
There are no further simple reductions.

In all examples, in this article, for a given $L$ we will exhibit
only the first non-trivial equation from the Lax hierarchy
\eqref{exev} .

\begin{example} $N=2$: The KP.

The Lax operator is $L=\prh_x^2+u$. Solving \eqref{exnov} for
$n=3$ one finds $(A_3)_{\me 0} =
\prh_x^3+\frac{3}{2}u\prh_x+\frac{3}{4}u_x+\frac{3}{4}\pr_x^{-1}u_y$.
Hence
\begin{equation}\label{kp}
u_{t_3} =
\frac{1}{4}u_{3x}+\frac{3}{2}uu_x+\frac{3}{4}\pr_x^{-1}u_{2y} .
\end{equation}
\end{example}
\begin{example} $N=3$: The (2+1)-Boussinesq.

Let $L=\prh_x^3+u\prh_x+v$. Then, for $(A_2)_{\me
0}=\prh_x^2+\frac{2}{3}u$
\begin{equation}\label{b}
    \pmatrx{u\\ v}_{t_2} = \pmatrx{2v_x-u_{2x}\\
    \frac{2}{3}u_y-\frac{2}{3}uu_x+v_{2x}-\frac{2}{3}u_{3x}} .
\end{equation}
\end{example}

Considering the restriction of the Lax operators to the form
\eqref{exl} we find
\begin{align}
\label{ek0} L &= \prh_x^N + u_{N-2} \prh_x^{N-2} + ... +  u_0 +
\sum_i \psi_i \prh_x^{-1} \phi_i & N\me 1,\\
\label{ek0N0} L &= \sum_i \psi_i \prh_x^{-1} \phi_i & N=0 .
\end{align}
The Lax operator \eqref{ek0N0} is pure (2+1)-dimensional effect
and does not have (1+1)-dimensional counterpart.

\begin{example} $N=1$: The (2+1) AKNS.

The Lax operator is $L = \prh_x + \psi \prh_x^{-1} \phi$. Then,
$(A_2)_{\me 0} = \prh_x^2 + a$ and
\begin{equation}\label{a}
\pmatrx{\psi\\ \phi}_{t_2} = \pmatrx{a \psi + \psi_{2x}\\ -a \phi
- \phi_{2x}}\qquad a_x-a_y-2\bra{\psi \phi}_x=0.
\end{equation}
\end{example}
\begin{example} $N=0$: Two-field system.

Let $L = \psi \prh_x^{-1} \phi$ and $(A_2)_{\me 0} = \prh_x^2
-2\pr_y^{-1}(\psi \phi)_x$. Then
\begin{equation}\label{a0}
\pmatrx{\psi\\ \phi}_{t_2} = \pmatrx{\psi_{2x}-2\psi \pr_y^{-1}(\psi \phi)_x\\
 - \phi_{2x}+2\phi \pr_y^{-1}(\psi \phi)_x} .
\end{equation}
\end{example}

\paragraph{The case: $k=1$.}\hfill \\

The simplest restrictions of Lax operators are given in the
following form
\begin{align}
\label{k1} L &= \prh_x^N + u_{N-1} \prh_x^{N-1} + u_{N-2}
\prh_x^{N-2} + ... + u_0 + \prh_x^{-1}u_{-1} & N\me 1,\\
\label{k1N0} L &= u_0 + \prh_x^{-1}u_{-1} & N=0.
\end{align}
 There are further admissible reductions given
by $\{u_{-1}=0\}$ and $\{u_{-1}=u_0=0\}$. The case \eqref{k1N0}
does not exist in (1+1) dimensions.

\begin{example} $N=2$: The mKP.

For $L = \prh_x^2+2u\prh_x$ and for $(A_3)_{\me 1} = \prh_x^3 +
3u\prh_x^2+\frac{3}{2} (u^2+u_x+\pr_x^{-1}u_y)\prh_x$ we find
\begin{equation}\label{mkp}
u_{t_3} =
\frac{1}{4}u_{3x}-\frac{3}{2}u^2u_x+\frac{3}{2}u_x\pr_x^{-1}u_y
+\frac{3}{4}\pr_x^{-1}u_{2y} .
\end{equation}
\end{example}
\begin{example} $N=3$: The (2+1) modified Boussinesq.

Let $L=\prh_x^3+3u\prh_x^2+v\prh_x$. Then, for $(A_2)_{\me 1} =
\prh_x^2+2u\prh_x$
\begin{equation}\label{mb}
 \pmatrx{u\\ v}_{t_2} = \pmatrx{\frac{2}{3}v_x-2uu_x-u_{2x}\\
 2u_y-2u_xv+2uv_x-6uu_{2x}+v_{2x}-2u_{3x}} .
\end{equation}
\end{example}
\begin{example} $N=1$: The (2+1) Kaup-Broer.

The Lax operator is $L=\prh_x + u + \prh_x^{-1}v$. For $(A_2)_{\me
1} = \prh_x^2 + a\prh_x$ we find
\begin{equation}\label{kb}
\pmatrx{u\\ v}_{t_2} = \pmatrx{2v_x+a u_x+u_{2x}\\
a_xv+av_x-v_{2x}}\qquad a_x-a_y-2u_x=0 .
\end{equation}
The reduction $v=0$ leads to the (2+1) Burgers equation: $u_{t_2}
= u_{2x}+ a u_x$ for $a$ given above.
\end{example}
\begin{example} $N=0$: Two-field system.

For \eqref{k1N0}: $L=u+\prh_x^{-1}v$ we construct $(A_2)_{\me 1} =
\prh_x^2-2\pr_y^{-1}u_x\prh_x$. Hence
\begin{equation}\label{kb0}
\pmatrx{u\\ v}_{t_2} = \pmatrx{2v_x+u_{2x}-2u_x\pr_y^{-1}u_x\\
-v_{2x}-2\bra{v\pr_y^{-1}u_x}_x} .
\end{equation}
Reducing it by the constraint $v=0$ one obtains a one-field system
${u}_{t_2} = u_{2x}-2u_x\pr_y^{-1}u_x$, which, according to the
transformation $u' = -\pr_y^{-1}u_x$, becomes the
(1+1)-dimensional Burgers equation $u'_t = u'_{2x} + 2u'u'_x$.
\end{example}

The admissible Lax operators \eqref{exl}, containing sources, have the form
\begin{align}
\label{ek1} L &= \prh_x^N + u_{N-1} \prh_x^{N-1} + ... + u_0 +
\prh_x^{-1}u_{-1} + \sum_i \psi_i \prh_x^{-1} \phi_i & N\me 1,\\
\label{ek1N0} L &= u_0 + \prh_x^{-1}u_{-1} + \sum_i \psi_i
\prh_x^{-1} \phi_i & N=0,
\end{align}
where \eqref{ek1N0} is pure (2+1)-dimensional phenomenon. There is
only one further reduction $\{u_{-1}=0\}$.

\begin{example} $N=1$: Three field system.

Let $L=\prh_x + u + \psi\prh_x^{-1}\phi$. Then, for $(A_2)_{\me 1}
= \prh_x^2 + a\prh_x$ we find
\begin{equation*}
\pmatrx{u\\ \psi\\ \phi}_{t_2} = \pmatrx{au_x+2(\psi \phi)_x +
u_{2x}\\ a\psi_x + \psi_{2x}\\ (a\phi)_x - \phi_{2x}}\qquad
a_x-a_y-2u_x=0.
\end{equation*}
\end{example}
\begin{example} $N=0$: Three field system.

For $L= u + \psi\prh_x^{-1}\phi$ and $(A_2)_{\me 1} = \prh_x^2 -
2\pr_y^{-1}u_x\prh_x$ we have
\begin{equation*}
\pmatrx{u\\ \psi\\ \phi}_{t_2} = \pmatrx{u_{2x}+2(\psi \phi)_x-2u_x\pr_y^{-1}u_x\\
\psi_{2x}-2\psi_x\pr_y^{-1}u_x\\ -
\phi_{2x}-2(\phi\pr_y^{-1}u_x)_x} .
\end{equation*}
\end{example}

We will show now, what will be important later, that fields $u_0$
and $u_{-1}$ from the Lax operators \eqreff{k1}{k1N0} and
\eqreff{ek1}{ek1N0} are expressible by the respective
eigenfunction and adjoint-eigenfunction in the following way
\begin{align}
\label{phi1} u_0 &= \varphi -\pr_x^{-1}u_{-1} & \text{for \eqreff{k1}{k1N0}},\\
\label{phi2} u_0 &= \varphi -\pr_x^{-1}u_{-1} - \sum_i
\psi_i\pr_x^{-1}\phi_i & \text{for \eqreff{ek1}{ek1N0}},
\end{align}
where $\varphi$ is an eigenfunction of the Lax hierarchy
\eqref{exev} for $k=1$ and $\psi_i$, $\phi_i$ are the source
fields. The following relations, fulfilled for an arbitrary operator
$A$ and an arbitrary smooth function $v$, will be useful later:
\begin{align}
\label{r1} & v\brac{A}_0 = \brac{vA}_0,\\
\label{r2} & \brac{A\prh_x^{-1}v}_0 = \brac{A\pr_x^{-1}v}_0 - \brac{A}_0\pr_x^{-1}v,\\
\label{r3} & \brac{\prh_x^{-1}A}_0 = \pr_x^{-1}\brac{A}_0 - \pr_x^{-1}\brac{A^\dag}_0 .
\end{align}
It is enough to consider the case \eqref{phi2} as the case
\eqref{phi1} is a simple consequence of \eqref{phi2}. Let $L = ...
+ u_0 + \prh_x^{-1}u_{-1} + \sum_i \psi_i \prh_x^{-1} \phi_i$,
then immediately from \eqref{trexev} it follows that
$(u_{-1})_{t_n} = - [(A)_{\me 1}^\dag u_{-1}]_0$. Hence, the field
$u_{-1}$ is an adjoint-eigenfunction. The time evolution for $u_0$
is obtained from \eqref{exev} as
\begin{equation}\label{evu0}
(u_0)_{t_n} = \brac{Bu_0}_0 + \brac{B\prh_x^{-1}u_{-1}}_0 -
\brac{\prh_x^{-1}u_{-1}B}_0 + \sum_i
\brac{B\psi_i\prh_x^{-1}\phi_i}_0 - \sum_i
\brac{\psi_i\prh_x^{-1}\phi_iB}_0,
\end{equation}
where $B=(A_n)_{\me 1}$. We introduce now a new function $\varphi$
defined as $\varphi = u_0 + \pr_x^{-1}u_{-1} + \sum_i \psi_i
\pr_x^{-1} \phi_i$ for which from \eqref{evu0} we obtain the
following time evolution
\begin{align*}
\varphi_{t_n} &= \brac{B\varphi}_0 - \brac{B\pr_x^{-1}\varphi}_0 - \sum_i \brac{B\psi_i\pr_x^{-1}\phi_i}_0 + \brac{B\psi_i\prh_x^{-1}u_{-1}}_0\\
&\quad - \brac{\prh_x^{-1}u_{-1}B}_0 + \sum_i \brac{B\psi_i\prh_x^{-1}\phi_i}_0 - \sum_i \brac{\psi_i\prh_x^{-1}\phi_iB}_0\\
&\quad - \pr_x^{-1}\brac{B^\dag u_{-1}}_0 + \sum_i \brac{B\psi_i}_0\pr_x^{-1}\phi_i - \sum_i \psi_i\pr_x^{-1}\brac{B^\dag \phi_i}_0 .
\end{align*}
As the following relations are valid:
\begin{align*}
&\brac{B\prh_x^{-1}u_{-1}}_0 = \brac{B\pr_x^{-1}v}_0 & \text{by \eqref{r2}},\\
&\brac{\prh_x^{-1}u_{-1}B}_0 = - \pr_x^{-1}\brac{B^\dag u_{-1}}_0 & \text{by \eqref{r3}},\\
&\brac{B\psi_i\prh_x^{-1}\phi_i}_0 = \brac{B\psi_i\pr_x^{-1}\phi_i}_0 - \brac{B\psi_i}_0\pr_x^{-1}\phi_i & \text{by \eqref{r2}},\\
&\brac{\psi_i\prh_x^{-1}\phi_iB}_0 = - \psi_i\pr_x^{-1}\brac{B^\dag \phi_i}_0 & \text{by \eqref{r1} and \eqref{r3}},
\end{align*}
time evolution of $\varphi$ is $\varphi_{t_n} = [(A_n)_{\me
1}\varphi]_0$. Hence, the field $\varphi$ is indeed an
eigenfunction of \eqref{exev} for $k=1$.

\paragraph{The case: $k=2$.}\hfill \\

Appropriate Lax operators are of the form
\begin{equation}\label{k2}
L = u_N \prh_x^N + u_{N-1} \prh_x^{N-1} + ... + u_{1} \prh_x + u_0
+ \prh_x^{-1}u_{-1} + \prh_x^{-2}u_{-2}\qquad N\me 1
\end{equation}
and simplest admissible reductions are given by $\{u_{-2}=0\}$,
$\{u_{-2}=u_{-1}=0\}$, $\{u_{-2}=u_{-1}=u_0=0\}$,
$\{u_{-2}=u_{-1}=u_0=u_1=0\}$.

\begin{example} $N=2$: The (2+1) HD.

For the Lax operator $L = u^2\prh_x^2$ and $(A_3)_{\me 2} =
u^3\prh_x^3 +
\frac{3}{2}u^2(u_x+\pr_x^{-1}\frac{u_y}{u^2})\prh_x^2$ we find
\begin{equation}\label{hd}
u_{t_3} =
\frac{1}{4}u^3u_{3x}+\frac{3}{4}\frac{1}{u}\bra{u^2\pr_x^{-1}\frac{u_y}{u^2}}_y
.
\end{equation}
\end{example}
\begin{example} $N=3$: Two-field system.

Let $L=u^3\prh_x^3+v\prh_x^2$ and $(A_2)_{\me 2} = u^2\prh_x^2$.
Then
\begin{equation}\label{s}
 \pmatrx{u\\ v}_{t_2} =
 \pmatrx{\frac{2}{3}v_x-\frac{4}{3}\frac{u_x}{u}v-u^2u_{2x}\\
 2uu_y-2u_x^2v-2uu_{2x}v+u^2v_{2x}-6u^3u_xu_{2x}-2u^4u_{3x} } .
\end{equation}
\end{example}
\begin{example} $N=1$: Three-field system.

For $L=u\prh_x+v+\prh_x^{-1}w$ and $(A_2)_{\me 2} = a\prh_x^2$ one
finds
\begin{equation}\label{s3}
 \pmatrx{u\\ v\\ w}_{t_2} =
 \pmatrx{2av_x+au_{2x}\\ a_xw+2aw_x+av_{2x}\\ -(aw)_{2x}}\qquad a_y-ua_x+2au_x=0 .
\end{equation}
\end{example}

Considering the restrictions \eqref{exl} one find that the appropriate
Lax operators are given by
\begin{equation}\label{ek2}
L = u_N \prh_x^N + u_{N-1} \prh_x^{N-1} + ... + u_0 +
\prh_x^{-1}u_{-1} + \prh_x^{-2}u_{-2} + \sum_i \psi_i \prh_x^{-1}
\phi_i \qquad N\me 1.
\end{equation}
The reductions are $\{u_{-2}=0\}$, $\{u_{-2}=u_{-1}=0\}$.

\section{Gauge transformation and reciprocal link}

The three classes of Lax hierarchies \eqref{exev} for $k=0, 1, 2$
are interrelated as shown in the following two theorems.

\begin{theorem}\label{gauge}
Gauge transformation: $k=0 \rightarrow k=1$. Let $L\in \cea$
satisfy the hierarchy $L_{t_n} = [(A_n)_{\me 0},L - \prh_y]$ and
let the function $\psi \neq 0$ be an eigenfunction of this
hierarchy: $\psi_{t_n} = [\bra{A_n}_{\me 0}\psi]_0$. Then,
$\tilde{L} = \psi^{-1} L \psi - \psi^{-1} \psi_y$ satisfies the
hierarchy $\tilde{L}_{t_n} = [(\tilde{A}_n)_{\me 1},\tilde{L} -
\prh_y]$, where $\tilde{A}_n = \psi^{-1} A_n \psi$.
\end{theorem}
\begin{proof}
First we have to show that $\tilde{A}_n$ is solution of
\eqref{exnov} for $\tilde{L}$:
\begin{flalign*}
\brac{\tilde{A}_n,\tilde{L}-\prh_y} &= \brac{\psi^{-1} A_n \psi,
\psi^{-1} L
\psi -\psi^{-1}\psi_y-\prh_y}\\
& = \psi^{-1} \brac{A_n, L - \psi^{-1}\psi_y - \psi \prh_y
\psi^{-1}} \psi = \psi^{-1} \brac{A_n, L  - \prh_y}\psi = 0 .
\end{flalign*}
Next, one observes that for an arbitrary pseudo-differential
operator $A$, the following relation is valid: $(\psi^{-1} A
\psi)_{\geq 1} = \psi^{-1} (A)_{\me 0} \psi - \psi^{-1} [(A)_{\me
0} \psi]_0$. Then
\begin{flalign*}
\brac{\bra{\tilde{A}_n}_{\me 1},\tilde{L} - \prh_y} & =
\brac{\psi^{-1} \bra{A_n}_{\me 0} \psi - \psi^{-1}
\brac{\bra{A_n}_{\me 0} \psi}_0, \psi^{-1} L
\psi -\psi^{-1}\psi_y-\prh_y}\\
& = \psi^{-1} \brac{(A_n)_{\me 0},L - \prh_y} \psi - \psi^{-1}
\brac{\psi^{-1} \brac{\bra{A_y}_{\me 0} \psi}_0, L - \prh_y} \psi.
\end{flalign*}
Now, since
\begin{equation*}
\tilde{L}_{t_n} = \bra{\psi^{-1} L \psi -\psi^{-1}\psi_y}_{t_n} =
\psi^{-1} L_{t_n} \psi - \psi^{-1} \brac{\psi^{-1} \psi_{t_n}, L -
\prh_y} \psi
\end{equation*}
the proof of the theorem is completed.
\end{proof}

Consider the Lax operator of the form \eqref{k0}. Then the gauge
transformed operator, by theorem \ref{gauge}, is given by
\begin{equation}\label{gk0}
  \tilde{L} = \prh_x^N + N\psi^{-1}\psi_x\prh_x^{N-1} + ... +
(...)\prh_x + u_0 + \psi^{-1}\brac{L\psi}_0 -\psi^{-1}\psi_y \qquad
  N\me 2.
\end{equation}
We will compare \eqref{gk0} with Lax operator \eqref{k1} with
$\tilde{u}_i$ components, where
$\{\tilde{u}_{-1}=\tilde{u}_0=0\}$, as then it is spanned by the
same number of dynamical fields as \eqref{k0}. Hence
\begin{align*}
  \tilde{u}_{N-1}&=N\psi^{-1}\psi_x\\
  \tilde{u}_{N-2}&=u_{N-1}+\frac{N(N-1)}{2}\psi^{-1}\psi_{2x}\\
&\ \vdots\\
0&=u_0 + \psi^{-1}\brac{L\psi}_0 -\psi^{-1}\psi_y .
\end{align*}
Now, eliminating eigenfunction $\psi$ one obtains the Miura
transformation between the fields $u_i$ from the Lax operator
\eqref{k0} for $k=0$ and the fields $\tilde{u}_i$ from the Lax
operator \eqref{k1} $\{\tilde{u}_{-1}=\tilde{u}_0=0\}$ for $k=1$.

\begin{example} A well known Miura map between systems KP \eqref{kp} and mKP
\eqref{mkp} is given in the form
\begin{equation*}
L=\prh_x^2+u\longrightarrow
\tilde{L}=\prh_x^2+2\tilde{u}\prh_x\quad \Longrightarrow \quad
    (u+\tilde{u}^2+\tilde{u}_x)_x=\tilde{u}_y .
\end{equation*}
\end{example}
\begin{example} The Miura map for (2+1) Boussinesq \eqref{b} and (2+1) modified Boussinesq
\eqref{mb} is
\begin{equation*}
L=\prh_x^3+u\prh_x+v\longrightarrow
\tilde{L}=\prh_x^3+3\tilde{u}\prh_x^2+\tilde{v}\prh_x\quad
\Longrightarrow \quad \alig{&u = \tilde{v}-3\tilde{u}^2-3\tilde{u}_x\\
&(v-2\tilde{u}^3+\tilde{u}\tilde{v}+\tilde{u}_{2x})_x=\tilde{u}_y
.}
\end{equation*}
\end{example}

Let us consider the source Lax operators given by the form
\eqref{ek0}. A natural choice of eigenfunctions is the choice of
one from the source eigenfunctions $\psi_i$. Let $\psi = \psi_1$,
as then the Lax operator \eqref{ek0} naturally transforms to the
\eqref{ek1} form. That is since the gauge transformed operator has
the form
\begin{equation}\label{gek0}
  \tilde{L} = \prh_x^N + N\psi^{-1}\psi_x\prh_x^{N-1} + ... +
u_0 + \psi^{-1}\brac{L\psi}_0 -\psi^{-1}\psi_y + \prh_x^{-1}\phi_1
+ \sum_{i\neq 1} \psi_i \prh_x^{-1} \phi_i\qquad
  N\me 1,
\end{equation}
then \eqref{ek0} and \eqref{ek1} are spanned by the same
number of dynamical fields. Analogously, if $\psi = \psi_1$ the
Lax operators of the form \eqref{ek0N0} naturally lead to the Lax
operators \eqref{ek1N0}. If operators \eqref{ek0} and
\eqref{ek0N0} contain only one pair of eigenfunction and
adjoint-eigenfunction $\psi_1, \phi_1$, then by theorem
\eqref{gauge} with $\psi=\psi_1$ we construct a Miura map between
the fields from the Lax operators \eqref{ek0}, \eqref{ek0N0} ($i\in
\{1\}$) and the Lax operators \eqref{k1}, \eqref{k1N0},
respectively.

\begin{example} The Miura map between (2+1) AKNS \eqref{a} and (2+1) KB
\eqref{kb} is given by
\begin{equation*}
L=\prh_x+\psi\prh_x^{-1}\phi\longrightarrow \tilde{L}=\prh_x+u +
\prh_x^{-1}v\quad
\Longrightarrow \quad \alig{u&=\psi^{-1}\psi_x-\psi^{-1}\psi_y\\
v&=\psi \phi.}
\end{equation*}
\end{example}
\begin{example} The transformation between fields for
two-field systems \eqref{a0} and \eqref{kb0} is
\begin{equation*}
L=\psi\prh_x^{-1}\phi\longrightarrow \tilde{L}= u +
\prh_x^{-1}v\quad
\Longrightarrow \quad \alig{u&=-\psi^{-1}\psi_y\\
v&=\psi \phi.}
\end{equation*}
\end{example}

\begin{theorem}\label{link}
The reciprocal link: $k=1 \rightarrow k=2$. Let $L=L(x,y,t)$ satisfy
$L_{t_n} = [(A_n)_{\me 1}, L - \prh_y]$ and the function
$\phi(x,y,t)$, such that $\varphi_x \neq 0$ and $\varphi_y \neq 0$, be
an eigenfunction of this hierarchy satisfying $\varphi_{t_n} =
[(A)_{\me 1}\varphi]_0$. Consider the following transformation
$x'=\varphi(x,y,t)$, $y'=y$, $t_n'=t_n$. Then,
$L'(x',y',t')=L(x,y,t)-\varphi_y \prh_{x'}$ satisfies the hierarchy
$L'_{t_n'} = [(A'_n)_{\me 2},L' - \prh_{y'}]$, where
$A_n'(x',y',t')= A_n(x,y,t)$.
\end{theorem}
\begin{proof}
Consider transformation: $x'=\varphi(x,y,t),\ y'=y,\ t_n'=t_n$, then
\begin{equation*}
  \pr_x = \varphi_x \pr_{x'}\quad \pr_y = \varphi_y \pr_{x'} + \pr_{y'},\quad
  \pr_{t_n} = \varphi_{t_n} \pr_{x'} + \pr_{t'_n}.
\end{equation*}
In consequence $L' - \prh_{y'} = L - \prh_{y}$ and $[A'_n,
L'-\prh_{y'}] = [A_n, L-\prh_{y}]=0$. Let $A=\sum_i
a_i\prh_{x'}^i$. Observing that for an arbitrary
pseudo-differential operator $(A')_{\me 1}$ the lowest coefficient
can be obtained by $a_1 = [(A')_{\me 1} x']_0$ one finds the
following relation $(A')_{\me 2} = (A)_{\me 1} - [(A)_{\me 1}
\varphi]_0 \prh_{x'}$. Hence
\begin{equation*}
\brac{\bra{A'_n}_{\me 2},L' - \prh_{y'}} = \brac{\bra{A_n}_{\me
1}, L-\prh_y} - \brac{\brac{\bra{A_n}_{\me 1} \varphi}_0 \prh_{x'}, L
- \prh_{y}} .
\end{equation*}
Now as
\begin{flalign*}
L'_{t'_n} &= \brac{\prh_{t'_n}, L'} =
\brac{\prh_{t_n}-\varphi_{t_n}\prh_{x'}, L -
\varphi_y \prh_{x'}}\\
&= L_{t_n} - \brac{\varphi_{t_n}\prh_{x'}, L-\prh_y}\\
&\qquad + \brac{ {\varphi_{t_n}}\bra{\varphi_x}^{-1}\prh_x,
{\varphi_y}\bra{\varphi_x}^{-1}\prh_x }- \brac{\prh_{t_n},
{\varphi_y}\bra{\varphi_x}^{-1}\prh_x} -
\brac{{\varphi_{t_n}}\bra{\varphi_x}^{-1}\prh_x , \prh_y},
\end{flalign*}
where the last three terms cancel to zero, we obtain the result of
the theorem.
\end{proof}

Consider the Lax operators \eqref{k1} with the reduction
$\{u_{-1}=u_0=0\}$. Then, the linked  operator from theorem
\ref{link}, where $x'=\varphi$, has the form
\begin{equation}\label{lk1}
    L' = \varphi_x^N\prh_{x'}^N +
    \brac{(N-1)\varphi_x^{N-2}\varphi_{2x}+u_{N-1}\varphi_x^{N-1}}\prh_{x'}^{N-1}
    + ... + \brac{...+u_1\varphi_x-\varphi_y}\prh_{x'}\qquad N\me 2 .
\end{equation}
We have to compare \eqref{lk1} with the Lax operators \eqref{k2}
where $\{u_{-2}=u_{-1}=u_0=u_{-1}=0\}$ as it is spanned by the
same number of dynamical fields as \eqref{k1} for
$\{u_{-1}=u_0=0\}$. So, it follows that the coefficient of
\eqref{lk1} standing by the first order term has to be equal zero.
Thus eliminating $\varphi$ we construct the reciprocal transformation.

\begin{example} A well known reciprocal link between systems mKP \eqref{mkp} and (2+1) HD
\eqref{hd} is
\begin{align*}
L=\prh_x^2+2u\prh_x\longrightarrow L'=u'^2\prh_{x'}^2 \quad
\Longrightarrow \quad (2uu'+u'_x)_x=u'_y\qquad x'=\pr_x^{-1}u'.
\end{align*}
\end{example}
\begin{example} The reciprocal link between (2+1) modified Boussinesq \eqref{mb} and
two field system \eqref{s} has the form
\begin{align*}
&L=\prh_x^3+3u\prh_x^2+v\prh_x\longrightarrow
L'=u'^3\prh_{x'}^3+v'\prh_{x'}^2 \quad \Longrightarrow\\
&\alig{&v'=3uu'^2+3u'u_x\\ &(vu'+3uu'_x+u'_{2x})_x=u'_y}\qquad
x'=\pr_x^{-1}u'.
\end{align*}
\end{example}

We will now consider the cases of Lax operators \eqref{k1} and
\eqref{k1N0} without reductions. Then, the linked operators for
eigenfunction $\varphi$ are:
\begin{align}
\notag L' &= \varphi_x^N\prh_{x'}^N +
\brac{(N-1)\varphi_x^{N-2}\varphi_{2x}+u_{N-1}\varphi_x^{N-1}}\prh_{x'}^{N-1} + ...\\
\label{gl1} &\qquad \qquad \qquad \qquad \qquad + \brac{...+u_1\varphi_x-\varphi_y}\prh_{x'}+u_0 +\prh_{x'}^{-1}\tfrac{u_{-1}}{\varphi_x} & N\me 2,\\
\label{gl2} L' &=  \bra{\varphi_x-\varphi_y}\prh_{x'} + u_0 + \prh_{x'}^{-1}\tfrac{u_{-1}}{\varphi_x} & N=1,\\
\label{gl3} L' &=  -\varphi_y\prh_{x'} + u_0 +
\prh_{x'}^{-1}\tfrac{u_{-1}}{\varphi_x} & N=0 .
\end{align}
Here the natural choice of eigenfunction $\varphi$ is given by
\eqref{phi1}: $\varphi = u_0 + \pr_x^{-1}u_{-1}$. We will compare
linked operators \eqreff{gl1}{gl3} with \eqref{k2}, for fields
$u_i'$, and reduction $\{u_{-2}'=0\}$. Thus, $u_0'=u_0$ and
$u_{-1}'=\tfrac{u_{-1}}{\varphi_x}$. Now, as
\begin{equation*}
\varphi_x = \varphi_x (u_0')_{x'} + u_{-1} \Longleftrightarrow
u_{-1} = \varphi_x\bra{1-(u_0')_{x'}}
\end{equation*}
we find constraint $u_{-1}' = \bra{1-(u_0')_{x'}}$. Hence, we
found new appropriate restriction for $k=2$:
\begin{equation}\label{nk2}
L' = u_N' \prh_{x'}^N + u_{N-1}' \prh_{x'}^{N-1} + ... + u_0' +\prh_{x'}^{-1} \bra{1-(u_0')_{x'}}\qquad \qquad N\me 1.
\end{equation}
Therefore, by theorem \ref{link}, we construct reciprocal links
for fields from Lax operators \eqreff{k1}{k1N0} to fields from
\eqref{nk2}. But operators \eqref{k1} for $N=1$ and \eqref{k1N0} are linked to the same Lax operator \eqref{nk2} with $N=1$.

\begin{example} The reciprocal link between (2+1) Kaup-Broer \eqref{kb} and the
system \eqref{s3} with the reduction $w=1-v_x$. Let
$L'=u'\prh_{x'}+v'+\prh_{x'}^{-1}(1-v'_{x'})$, then for
$(A_2)_{\me 2} = a\prh_{x'}^2$ one finds
\begin{equation}\label{s2}
 \pmatrx{u\\ v}_{t_2} =
 \pmatrx{2av'_{x'}+au'_{2x'}\\ a_{x'}w'-a_{x'}v'_{x'}-av'_{2x'}}
 \qquad a_{y'}-u'a_{x'}+2au'_{x'}=0 .
\end{equation}
The reciprocal link is:
\begin{align*}
&L=\prh_x+u+\prh_x^{-1}v\longrightarrow
L'=u'\prh_{x'}+v'+\prh_{x'}^{-1}(1-v'_{x'}) \quad \Longrightarrow\\
&\alig{u'_x&=u_{2x}-u_{xy}+v_x-v_y\\ v'&=u}\qquad
x'=u+\pr_x^{-1}v.
\end{align*}
\end{example}
\begin{example} The reciprocal link between the two-field system \eqref{kb0} and the
system \eqref{s2}:
\begin{align*}
&L=u+\prh_x^{-1}v\longrightarrow
L'=u'\prh_{x'}+v'+\prh_{x'}^{-1}(1-v'_{x'}) \quad \Longrightarrow\\
&\alig{u'_x&=-u_{xy}-v_y\\ v'&=u}\qquad x'=u+\pr_x^{-1}v.
\end{align*}
\end{example}

The linked Lax operators \eqreff{ek1}{ek1N0} have similar form as
\eqreff{gl1}{gl3} with additional source terms, so $L' = ... +
\sum_i \psi_i' \prh_{x'}^{-1} \tfrac{\phi_i}{\varphi_x}$. We
choose eigenfunction $\varphi$ as $\varphi = u_0 +
\pr_x^{-1}u_{-1} + \sum_i \psi_i\pr_x^{-1}\phi_i$ by \eqref{phi2}.
We will equate $L'$ to the Lax operators of the form \eqref{ek2},
for fields $u_i'$, with reduction $\{u_{-2}'=0\}$. Thus,
$u_0'=u_0$, $u_{-1}'=\tfrac{u_{-1}}{\varphi_x}$, $\psi_i'=\psi_i$
and $\phi_i'=\tfrac{\phi_i}{\varphi_x}$. Now, as
\begin{equation*}
\varphi_x = \varphi_x (u_0')_{x'} + u_{-1} + \sum_i \varphi_x
\bra{\psi_i'\pr_{x'}^{-1}\phi_i'}_{x'}\Longleftrightarrow u_{-1} =
\varphi_x\bra{1-(u_0')_{x'}-\sum_i
\bra{\psi_i'\pr_{x'}^{-1}\phi_i'}_{x'}}
\end{equation*}
we find constraint $u_{-1}' = 1-(u_0')_{x'}-\sum_i
\bra{\psi_i'\pr_{x'}^{-1}\phi_i'}_{x'}$. Hence, we found again new
restriction with sources for $k=2$ given by
\begin{equation}\label{enk2}
L' = u_N' \prh_{x'}^N + ... + u_0' +\prh_{x'}^{-1}
\bra{1-(u_0')_{x'}-\sum_i \bra{\psi_i'\pr_{x'}^{-1}\phi_i'}_{x'}}
+ \sum_i \psi_i'\prh_{x'}^{-1}\phi_i'\qquad N\me 1.
\end{equation}
Therefore, by theorem \ref{link}, we construct reciprocal links
for the fields from Lax operators \eqreff{ek1}{ek1N0} to those fields from
\eqref{enk2}, respectively.

\section{Summary}

In this paper, we have examined the restrictions of Lax operators
allowing systematic construction of integrable (2+1)-dimensional
systems from three classes of Lax hierarchies \eqref{exev} in
centrally extended PDO algebra. It is important to mention that
systems \eqref{exev} are Hamiltonian, i.e. we can construct,
beside infinite hierarchy of commuting symmetries, also Poisson
tensor and infinitely many conserved quantities (see
\cite{BS,BS2}). Then, invariants solving the Novikov-Lax equation
\eqref{exnov} are differentials of Casimir functionals of the
natural Lie-Poisson bracket.

Besides, we have established some relations between the three
classes of Lax hierarchies \eqref{exev}. Theorem \ref{gauge}
describes the gauge transformation from $k=0$ to $k=1$ allowing
a construction of Miura maps between respective evolution equations.
Theorem \ref{link} shows how to construct reciprocal links from
$k=1$ to $k=2$ for respective systems. It may be worth further
investigation to extend the presented theory to other gauge,
reciprocal and as well as B{\"a}cklund and auto-B{\"a}cklund
transformation for Lax hierarchies \eqref{exev}, in a similar way
as it is done for the (1+1)-dimensional case \cite{KO2,OR}.

\paragraph{Acknowledgment}\hfill \\

The author is grateful to M. B\l aszak for useful and stimulating
discussions. This work was partially supported by KBN research
grant No. 1 P03B 111 27.

\label{lastpage}


\begin{thebibliography}{99}
\small



\bibitem{BM} B\l aszak M and Marciniak K, R-matrix approach to lattice integrable
systems {\it J. Math. Phys.} {\bf 35} (1994) 4661


\bibitem{BS1} B\l aszak M and Szablikowski B M, Classical $R$-matrix theory of dispersionless systems: I. (1+1)-dimension theory {\it J. Phys A: Math. Gen.,} {\bf 35} (2002) 10325-44

\bibitem{BS2} B\l aszak M and Szablikowski B M, Classical $R$-matrix theory of dispersionless systems: II. (2+1) dimension theory {\it J. Phys A: Math. Gen.,} {\bf 35} (2002) 10345-64

\bibitem{BS3} B\l aszak M and Szablikowski B M, From dispersionless to soliton
systems via Weyl-Moyal-like deformations {\it J. Phys A: Math.
Gen.,} {\bf 36}  (2003) 12181-203

\bibitem{BS} B\l aszak M and Szum A, Lie algebraic approach to the construction
of (2+1)-dimensional lattice-field and field integrable
Hamiltonian equations {\it J. Math. Phys.} {\bf 35} (2001) 4088-116

\bibitem{BSP} B\l aszak M, Szum A and Prykarpatsky A, Central extension approach to integrable field and lattice-field systems in (2+1)-dimensions,  {\it Rep. Math. Phys.} {\bf 44} (2001) 37-44

\bibitem{KD} Konopelchenko B G and Dubrovsky V G, Some new integrable nonlinear
evolution equations in 2+1 dimensions {\it Phys. Lett. A} {\bf 102}
 (1984) 15-17

\bibitem{KO1} Konopelchenko B and Oevel W, Matrix Sato theory and integrable equations in 2+1 dimensions, in: {\it Nonlinear evolution equations and dynamical systems}, eds. M.Boiti, L.Martina and F.Pempinelli, World Scientific, Singapore, 1991

\bibitem{KO2} Konopelchenko B G and Oevel W, An r-matrix approach to nonstandard
 classes of integrable equations {\it Publ. RIMS, Kyoto Univ.} {\bf 29} (1993) 581-666

\bibitem{Li} Li L C, Classical r-Matrices and Compatible Poisson Structures
for Lax Equations in Poisson Algebras {\it Commun. Math. Phys.} {\bf 203} (1999) 573-92

\bibitem{OR} Oevel W and Rogers C, Gauge transformations and reciprocal links in
2+1 dimensions {\it Rev. Math. Phys.} {\bf 5} (1993) 299

\bibitem{OS} Oevel W and Strampp W, Constrained KP hierrachy and
bi-Hamitonian structures, {\it Commun. Math. Phys.} {\bf 157}
(1993) 51

\bibitem{PSA} Prykarpatsky A K, Samoilenko V Hr and Andrushkiw R I,
Algebraic structure of the gradient-holonomic algorithm for Lax
integrable nonlinear dynamical systems I {\it J. Math. Phys.} {\bf
35} (1994) 1763-77

\bibitem{S} Sato M, Soliton equations as dynamical systems on infinite Grassmann manifold, {\it RIMS Kokyuroku, Kyoto Univ.} {\bf 439} 30

\bibitem{S-T-S} Semenov-Tian-Shansky M A, What is a classical r-matrix?
  {\it Funct. Anal. Appl.} {\bf 17} (1983) 259

\bibitem{RS-T-S1} Reyman A G and Semenov-Tian-Shansky M A, Current algebras and nonlinear partial differential equations, (in russian) {\it Dokl. Akad. nauk} {\bf 251} 1310

\bibitem{RS-T-S2} Reyman A G and Semenov-Tian-Shansky M A, Hamiltonian structure
of Kadomtsev-Petviashvili type equations, (in russian)  {\it LOMI}
{\bf 133} (6) (1984) 212


\end{thebibliography}
\end{document}